\documentstyle[12pt,graphics]{article}
\def\reference{\parskip 0pt\par\noindent\hangindent 0.5 truecm}

\begin{document}

\title{Comprehensive analysis of $RGU$ photometry in the direction to M5}
\author{S.~Karaali,$^{1}$ 
 S.~Bilir,$^{1}$ 
 R.~Buser$^{2}$ \and
 } 
\date{}
\maketitle

{\center
$^1$ Istanbul University Science Faculty, 
Department of Astronomy and Space Sciences, 34452 University-
Istanbul, Turkey \\karsa@istanbul.edu.tr\\
                  sbilir@istanbul.edu.tr\\[3mm]
$^2$ Astronomisches Institut der Universit\"{a}t Basel, 
Venusstr. 7, 4102 Binningen, Switzerland\\Roland.Buser@unibas.ch\\[3mm]
}

\begin{abstract}

The $RGU$-photographic investigation of an intermediate latitude field in the 
direction to the Galactic center is presented. 164 extra-galactic objects, 
identified by comparison of Minnesota and Basel charts, are excluded from the 
program. Also, a region with size 0.104 square-degrees, contaminated by 
cluster (M5) stars and affected by background light of the bright star 
HD 136202 is omitted. Contrary to previous investigations, a reddening of 
$E(B-V)=0.046$, corresponding to $E(G-R)=0.07$ mag is adopted. The separation 
of dwarfs and evolved stars is carried out by an empirical method, already 
applied in some of our works. A new calibration for the metallicity 
determination is used for dwarfs, while the absolute magnitude determination 
for stars of all categories is performed using the procedures given in the
literature. There is good agreement between the observed logarithmic space 
density histograms and the galactic model gradients. Also, the local 
luminosity function agrees with Gliese's (1969) and Hipparcos' 
(Jahreiss \& Wielen 1997) luminosity functions, for stars with $2<M(G)\leq8$ 
mag. For giants, we obtained two different local space densities from 
comparison with two Galactic models, i.e. $D^{*}(0)=6.63$, close to that of 
Gliese (1969), and $D^{*}(0)=6.79$. A metallicity gradient, 
$d[Fe/H]/dz= -0.20$ dex/kpc, is detected for dwarfs (only) with absolute 
magnitudes $4<M(G)\leq6$, corresponding to a spectral type interval F5-K0.

\end{abstract}

{\bf Keywords: Galaxy: structure -- Galaxy: metallicity gradient --
     Stars: luminosity function}
\bigskip

\section{Introduction}

The photographic data presented in this paper have been obtained in the 
context of the re-evaluation program of the Basel $RGU$ three-colour 
photometric high-latitude survey of Galaxy, which comprises homogeneous 
magnitudes and colours for about 20000 stars in a total of fourteen fields 
distributed along the Galactic meridian through the Galactic center and 
the Sun (Buser \& Rong 1995). The main purpose of the present investigation 
of this field near M5 is to check the possible metallicity gradient, 
claimed by many researchers (cf.Reid \& Majewski 1993, Chiba \& Yoshii 1998). 
This can be carried out employing known solar neighbourhood constraints, 
especially with consistency of the local stellar luminosity function. Thus, 
all methodical tools have been used to obtain a stellar luminosity function 
agreeable with that of Gliese (1969) and Hipparcos (Jahreiss \& Wielen 1997), 
as explained in the following sections. In  Section 2, we describe the 
identification of extra-galactic objects, the contamination with cluster stars, 
the background effect of the bright star HD 136202, and the two-colour diagrams. 
Section 3 is devoted to the separation of evolved stars (sub-giants and giants) 
and dwarfs, and the determination of absolute magnitudes and metallicities. 
The evaluation of density and luminosity functions is given in Section 4. 
In Section 5, the metallicity distribution is discussed and in Section 6, 
we provide a summary and brief discussion.    

\section{The data}
\subsection{Extra-galactic objects, cluster stars, stars affected by 
background light, and stars absent on Minnesota charts}

The comparison of Basel and Minnesota charts revealed that there is a 
considerable number of extra-galactic objects in the star fields, which cause 
an excess in the density and luminosity functions (Bilir et al. 2003). Hence, 
we applied the same procedure to eliminate such objects in our field. It turned 
out that 164 sources are extra-galactic objects, i.e. quasars or galaxies, 
occupying different regions in the two-colour diagrams (Fig.1). All these 
objects have been excluded from the program. Also, comparison of the number of 
stars per square-degree in the vicinity of the cluster M5 ($l=4^{o}.0$, 
$b=+47^{o}.0$, size 1.05 square-degrees) and at relatively farther distances 
revealed that the field is contaminated by cluster stars, causing a similar 
effect as just cited above. Additionally, some stars are affected by background 
light of the bright star HD 136202 
($\alpha = 15^{h}~ 19^{m}~18^{s}.80$, $\delta=+01^{o}~45{'}~55{''}.5$) in our 
field. To avoid such an effect a region with size 0.104 square-degrees was 
excluded from the field (Fig.2). Finally 33 objects which do not appear on 
either the Basel or the Minnesota charts have been omitted. Hence, a total 
number of 1368 stars have been included in the analysis within the limiting 
apparent magnitude $G=18.5$ and within the field of size $0.954$ square-degrees.

\begin{figure}
\resizebox{15cm}{!}{\includegraphics*{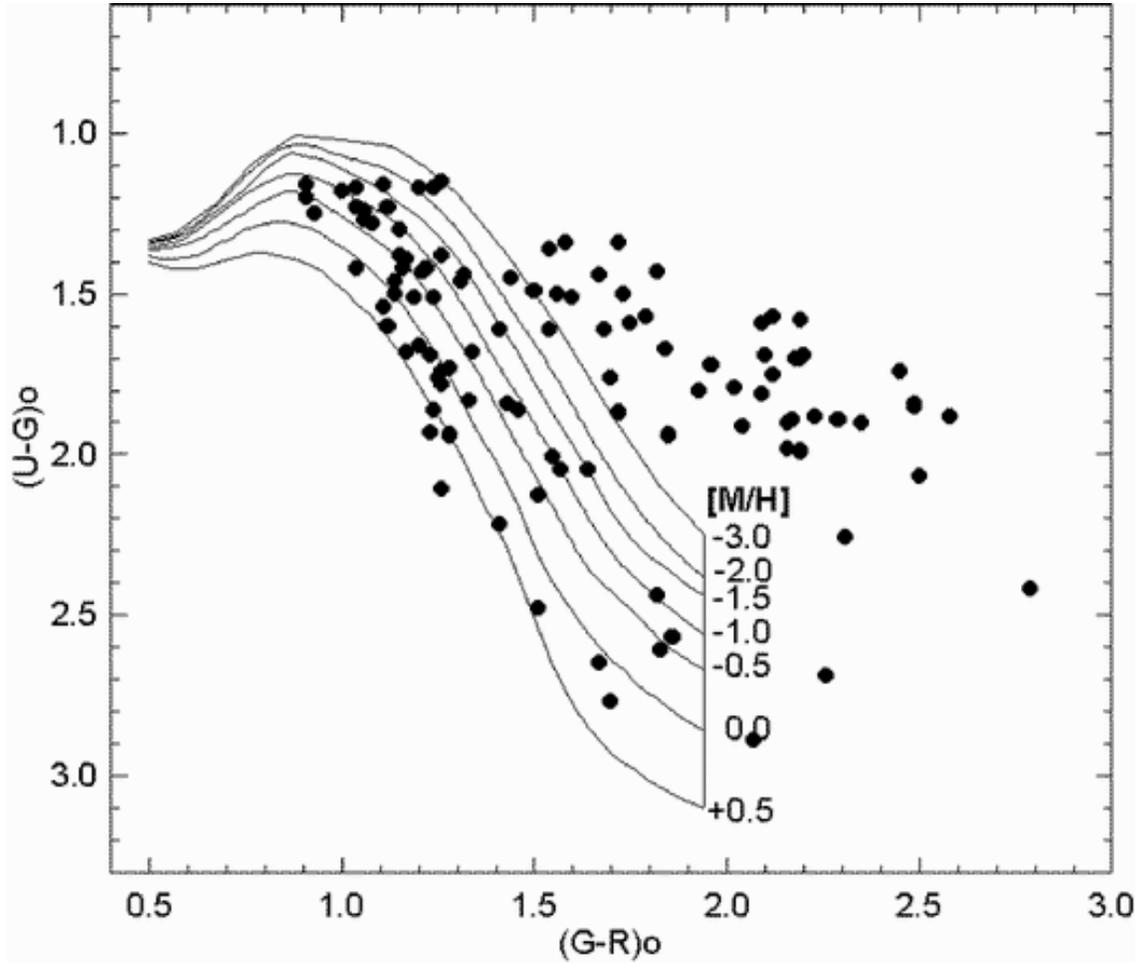}}
\caption{Extra-galactic objects identified by comparison of Minnesota and 
Basel charts. Their number within the limiting apparent magnitude, $G=18.5$, 
is 104, but increases up to 164 when counted down to the faintest object in 
the Basel catalogue.}
\end {figure}

\begin{figure}
\resizebox{15cm}{!}{\includegraphics*{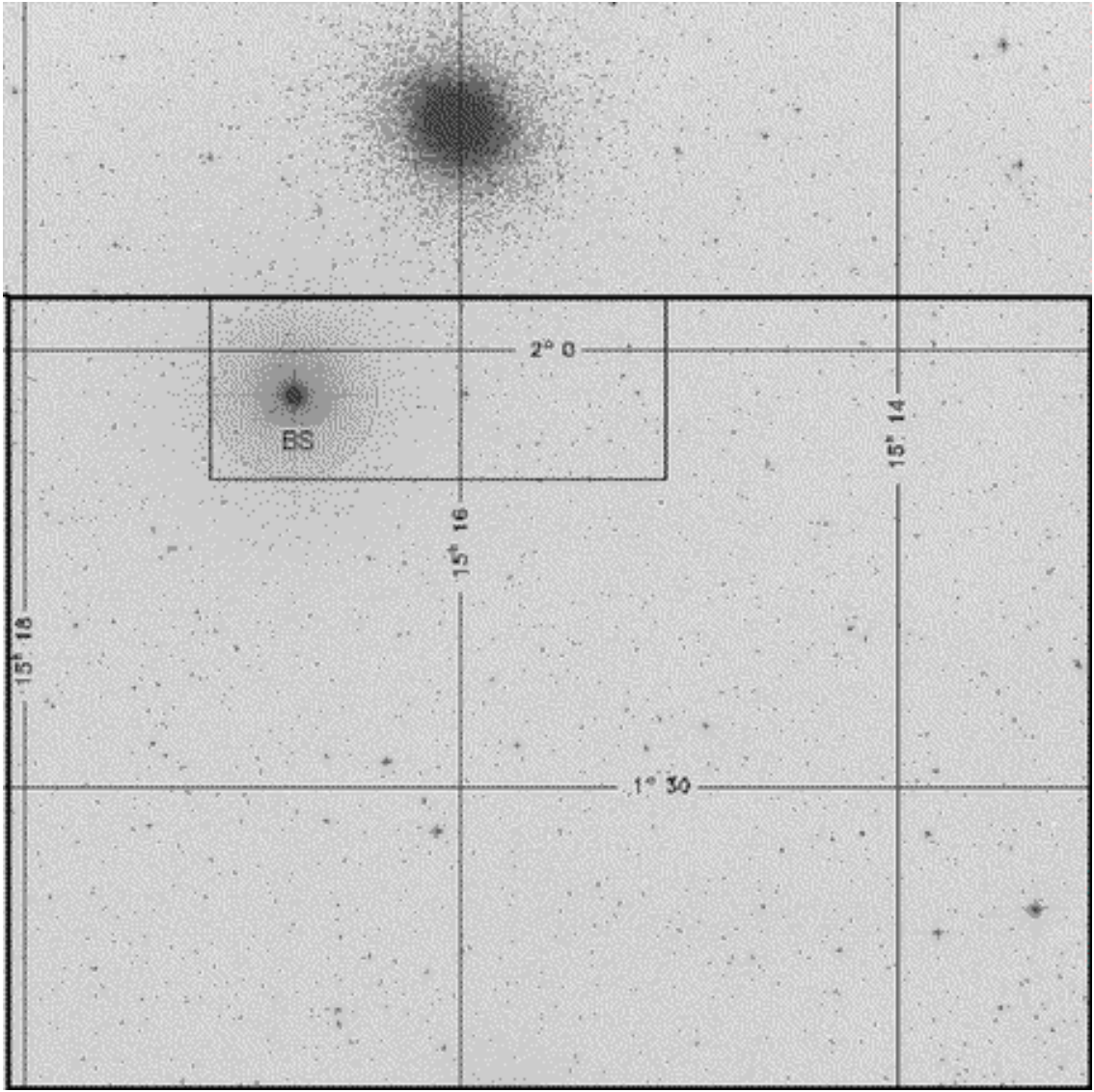}}
\caption{The field chart showing the position of the cluster M5, the bright 
star HD 136202 ($\alpha = 15^{h}~ 19^{m}~18^{s}.80$, 
$\delta=+01^{o}~45{'}~55{''}.5$), labelled with BS, and the surrounding 
excluded area of 0.104 square-degrees. The horizontal and vertical axes give 
the right ascension ($\alpha$) and declination ($\delta$), respectively.}
\end {figure}

\subsection{$RGU$ magnitudes, interstellar reddening, and two-colour diagrams}

The measurements, carried out by an automatic plate measuring machine (COSMOS) 
in 1980s in Edinburgh Royal Observatory, are transformed to the $RGU$-system 
according to Buser's (1978) formulae, with the help of 26 photoelectric 
$UBV$-standards. Although zero reddening was adopted in former investigations 
of this field (Becker et al. 1978, Fenkart \& Karaali 1990, Buser et al. 1998; 
1999), we adopted the $E(B-V)=0.046$ cited by Schlegel et al. (1998), which 
corresponds to $E(G-R)=0.07$ mag in Buser's system. The resulting extinction 
of this reddening is $A(G)=2.7E(G-R)=0.19$ mag. Thus, all colours and magnitudes 
used in this work are de-reddened. We fixed the limiting apparent magnitude at 
$G=18.5$, and omitted stars fainter than this magnitude (Fig.3).

\begin{figure}
\resizebox{15cm}{!}{\includegraphics*{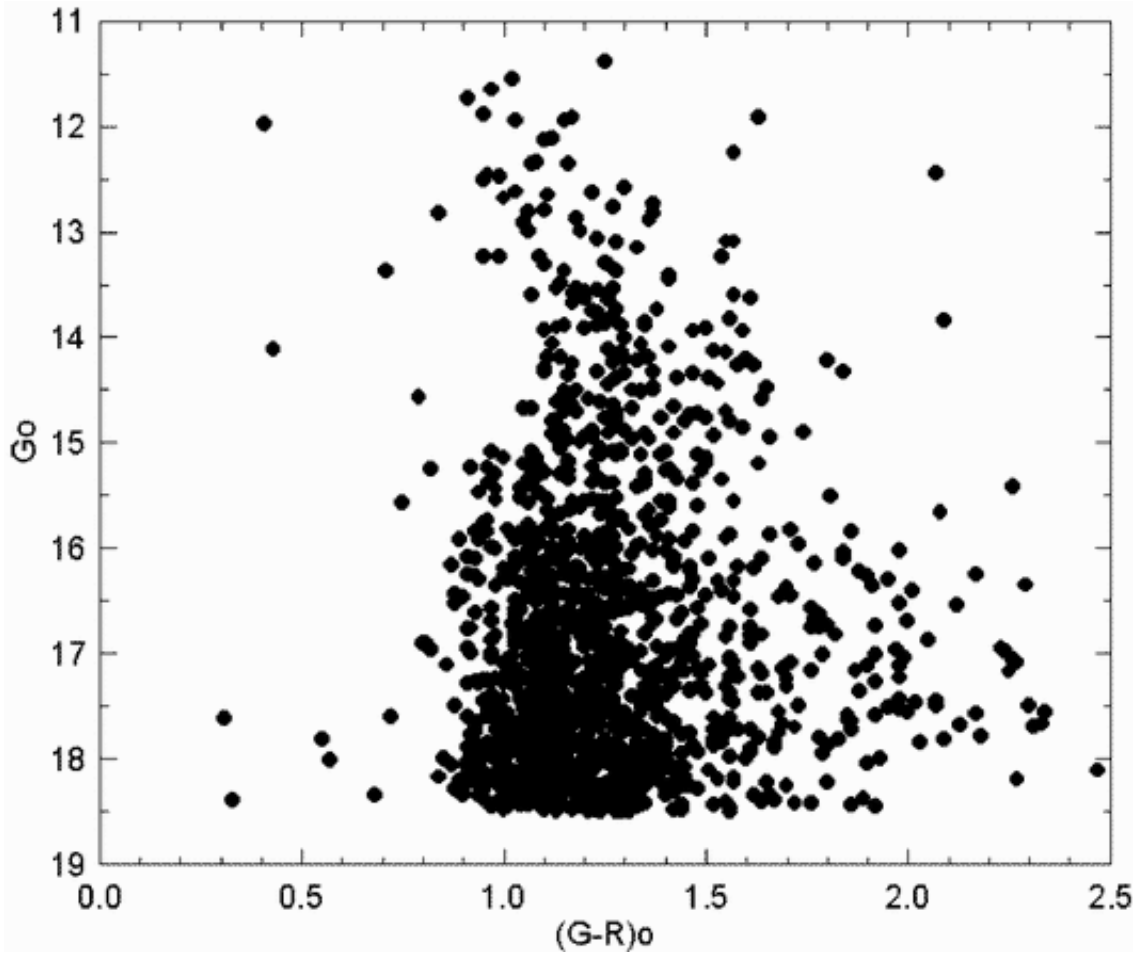}}
\caption{Colour-magnitude diagram for the field down to the limiting 
magnitude, $G=18.5$.}
\end {figure}

\begin{figure*}
\resizebox{15cm}{!}{\includegraphics*{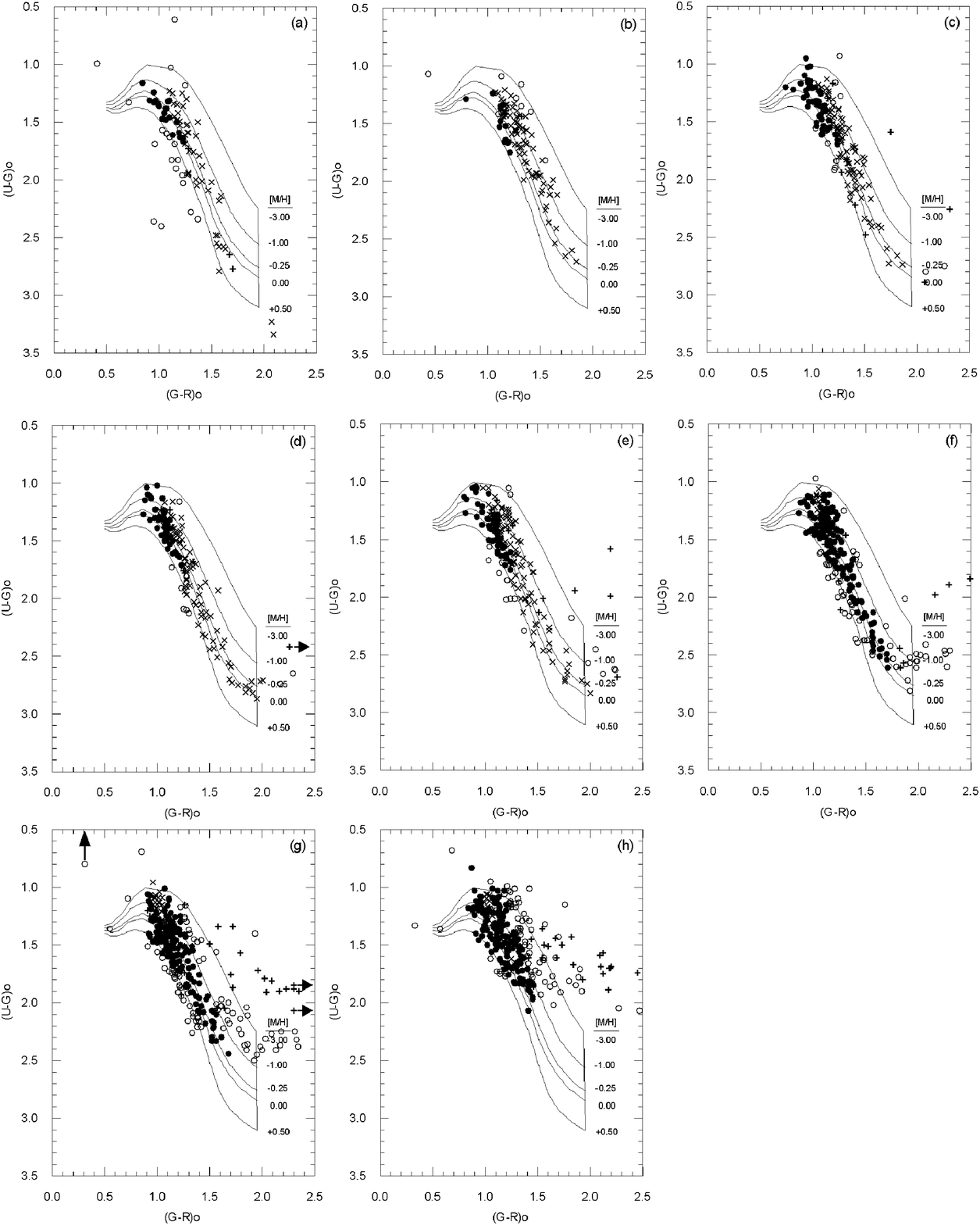}}
\caption{Two-colour diagrams in consecutive apparent magnitude intervals, 
i.e. (a) $G\leq14.0$, (b) $14.0<G\leq15.0$, (c) $15.0<G\leq16.0$, 
(d) $16.0<G\leq16.5$, (e) $16.5<G\leq17.0$, (f) $17.0<G\leq17.5$, 
(g) $17.5<G\leq18.0$, and (h) $18.0<G\leq18.5$. The superposed grid of 
iso-metallicity lines for dwarf stars is based on theoretical model 
atmosphere calculations (Buser \& Fenkart 1990). Symbols: ($\bullet$) dwarf, 
(x) evolved, (O) untreated, and (+) extra-galactic objects.}
\end{figure*}

The total number of stars in the sample is 1368. Their numbers in each panel 
on Fig.4 are given in Table 1 together with the numbers of extra-galactic objects.
The two-colour diagrams given in consecutive apparent magnitude intervals 
(Fig.4) is typical for an intermediate latitude field in the center direction 
of the Galaxy, i.e. most of the stars lie in the regions occupied by 
metal-rich or intermediately metal-rich stars, whereas the metal-poor stars 
are rare. Also, the scattering is less when compared with our recent works 
(cf. Karata\c{s} et al. 2001). Though, there are 262 stars which occupy the 
metallicity regions $[Fe/H]>+0.5$ dex or $[Fe/H]<-3$ dex where, usually, stars 
do not exist or they are rather sparse. Hence, these stars were excluded from 
the program without any inquiring. However, most of them (206) are relatively 
faint ones, $G>17.0$ mag, therefore they may be undetected blended stars. 
The exclusion of these extreme stars do not affect the metallicity distribution 
(see Section 6). Additionally, as they lie in a large range of the colour-index 
$(G-R)_{o}$ almost uniformly (Fig.4), they attribute to different $M(G)$ 
absolute magnitudes. Hence, they do not affect the luminosity function 
(see Section 4, Fig.7) either.   

\begin{table}
\caption{Number of stars and extra-galactic objects within the limiting 
magnitude $G=18.5$ for different panels in Fig.4.}
\begin{center}
{\scriptsize
\begin{tabular}{lrcc}
\hline
Panel & Apparent  & Number of & Number of \\
& magnitude & stars & extra-galactic objects \\
\hline
Fig.4a & $~~~~~~~~~~G\leq14.0$&~85& ~3 \\
Fig.4b & $14.0 < G\leq15.0$ & ~96 & ~3 \\
Fig.4c~& $15.0 < G\leq16.0$ & 154 & 11 \\
Fig.4d & $16.0 < G\leq16.5$ & 141 & ~7 \\
Fig.4e~& $16.5 < G\leq17.0$ & 157 & 11 \\
Fig.4f~& $17.0 < G\leq17.5$ & 220 & 13 \\
Fig.4g & $17.5 < G\leq18.0$ & 258 & 28 \\
Fig.4h & $18.0 < G\leq18.5$ & 257 & 28 \\
\hline
           &             Total  &1368 &104 \\
\hline
\end{tabular}
}  
\end{center}
\end{table}

\section{Separation of evolved stars, metallicity and absolute magnitude  
determination}

Following our recent experiences (Karaali 1992, Karaali et al. 1997, Ak et al. 
1998 Karata\c{s} et al. 2001, and Karaali et al. 2003), for apparent magnitudes 
brighter than $G=17$, stars which according to their positions in the two-colour 
diagram could be identified as dwarfs with assigned absolute magnitudes fainter 
than $M(G)=6$ (but see Section 4), are most likely evolved sub-giant or giant 
stars with correspondingly brighter absolute magnitudes, and their metallicities 
and absolute magnitudes are determined by the procedure given by Buser et al. 
(2000). Metallicities for dwarfs were determined using a new calibration, similar 
to that of Carney (1979), i.e. $[Fe/H]=0.11-2.22\delta-7.95\delta^{2}$, where 
$\delta$ is the ultra-violet excess at $G-R=1.08$ mag, corresponding to 
$B-V=0.60$ mag (Karaali \& Bilir 2002, see Appendix), and their absolute 
magnitudes are determined by means of the colour-magnitude diagrams of Buser \& 
Fenkart (1990). The scale of the new metallicity calibration, 
$-2.20\leq[Fe/H]\leq+0.20$ dex, is large enough to cover most of the dwarfs in 
our field. Actually, the number of dwarfs whose metallicities lie out of this 
interval is not large, so does not affect the metallicity distribution (see 
Section 5). 

\section {Density and luminosity functions}

The logarithmic space densities $D^{*}=logD(r)+10$ for stars of all population 
types, i.e. Population I (thin disk), Intermediate Population II (thick disk),
and Extreme Population II (halo) are given in Tables 2 and 3, for dwarfs and 
sub-giants, and giants, respectively. Here, $D=N/\Delta V_{1,2}$, N being the 
number of stars, found in the partial volume $\Delta V_{1,2}$, which is 
determined by the limiting distances $r_{1}$ and $r_{2}$, and apparent field 
size in square degrees $\Box$, i.e. 
$\Delta V_{1,2}=(\pi/180)^{2}(\Box/3)(r_{2}^{3}-r_{1}^{3})$. As usual, density 
functions are then given in the form of histograms with density plotted as a 
solid dot at the centroid distance, $\bar{r}=[(r^{3}_{1}+r^{3}_{2})/2]^{1/3}$,
of the corresponding volume, $\Delta V_{1,2}$ (see, e.g., Del Rio \& Fenkart 
1987, and Fenkart \& Karaali 1987). 

The comparison of the density functions with the best fitting model gradients 
predicted by Buser, Rong, \& Karaali (BRK 1998, 1999) and by Gilmore \& Wyse 
(GW 1985) are matched to the observed profiles in order to extrapolate the 
local stellar space densities. In both works the model gradients for thin disk 
and thick disk are calculated from the double exponentials fitted to them, 
whereas for halo the de Vaucouleurs spheroid is used for this purpose. The 
model gradients compared with the observed density functions are the combined 
ones for three galactic components, i.e. thin disk, thick disk, and halo. 
A small disagreement between the observed data and both models was noticed 
only for two absolute magnitude intervals, i.e. the excess number of stars 
with $5<M(G)\leq6$ within the distance interval $1.59<r\leq 3.98$ kpc and the 
deficient number of stars with $3<M(G)\leq4$ beyond the distance $r=3.98$ kpc. 
Assuming that about 60 stars in the fainter absolute magnitude interval are 
evolved leads to observed densities near the predicted model gradients for both 
absolute magnitude intervals, i.e. $3<M(G)\leq4$ and $5<M(G)\leq6$. Most of 
the stars cited above turned out to be with absolute magnitudes $3<M(G)\leq4$, 
and about a dozen of them with $4<M(G)\leq5$ mag. Their apparent magnitudes 
are fainter than $G=16$, however the peak of their magnitude distribution lies 
at $G\cong17.5$ mag Tables 2 and 3 give the final results.

Thus, we obtain good agreement between the gradients for both models, BRK and 
GW, and the observed logarithmic space density histograms (Fig.5 and Fig.6). 
The same holds also for the local densities, except for the giants, as 
explained as follows: the stellar luminosity function resulting from comparison
of observed histograms with the best fitting BRK- and GW-model gradients agrees
with the Gliese's (1969) and Hipparcos' (JW 1997) luminosity functions for all 
absolute magnitude intervals, i.e. $2<M(G)\leq3$, $3<M(G)\leq4$, $4<M(G)\leq5$, 
$5<M(G)\leq6$, $6<M(G)\leq7$, and $7 <M(G)\leq8$ (Fig.7). However, two different 
local densities are obtained for giants. Comparison with the GW model can be 
carried out up to the limiting distance of completeness, $r=15.85$ kpc, 
corresponding to a height of $z=11.59$ kpc from the Galactic plane, and gives a 
local density of $D^{*}(0)=6.63$, rather close to Gliese's (1969) value, 
$\odot=6.64$. At this distance the observed density falls abruptly and  
diverges from these model gradients for larger distances. On the other hand, 
for the model gradients from BRK, the comparison up to $r=15.85$ kpc gives a 
local density $D^{*}(0)= 6.79$, and the agreement holds up to the distance 
$r=19.95$ kpc with a local density slightly different from the previous one, 
$D^{*}(0)=6.77$.

\begin{table*}
\caption{The logarithmic space densities $D^{*}=logD+10$ for dwarfs and 
sub-giants of all population types, where $D=N/\Delta V_{1,2}$, N being the 
number of stars found in the partial volume $\Delta V_{1,2}$, determined by 
the limiting distances $r_{1}$ and $r_{2}$, and apparent field size in square 
degrees $\Box$, i.e. $\Delta V_{1,2}=(\pi/180)^{2}(\Box/3)(r_{2}^{3}-r_{1}^{3})$. 
$\bar{r}$: centroid distance $\bar{r}=[(r^{3}_{1}+r^{3}_{2})/2]^{1/3}$,
of the corresponding volume, $\Delta V_{1,2}$, heavy horizontal lines: limiting 
distance of completeness (distances in kpc, volumes in pc$^{3}$).}
\begin{center}
{\scriptsize
\begin{tabular}{ccccccccc}
\hline
 &    &    $M(G)\rightarrow$& (2-3]  & (3-4]    & (4-5]    & (5-6] & (6-7] & (7-8]\\
\hline
$r_{1}$-$r_{2}$   & $\Delta V_{1,2}$ & $\bar{r}$&  N~~D*   &  N~~D* &  N~~D* & N~~D* & N~~D*  & N~~D*\\
\hline
0.00-1.00 & 9.69 (4) & 0.79 &         &          &          &  34~~6.55 &         & 23~~6.38\\
0.00-1.59 & 3.86 (5) & 1.26 & ~7~~5.26& 28~~5.86 &  32~~5.92&           & 45~~6.07&         \\
1.00-1.26 & 9.64 (4) & 1.14 &         &          &          &           &         & \underline {22}~~\underline {6.36}\\
1.00-1.59 & 2.89 (5) & 1.36 &         &          &          &  57~~6.30 &         &         \\
1.26-1.59 & 1.92 (5) & 1.44 &         &          &          &           &         & 21~~6.04\\
1.59-2.00 & 3.83 (5) & 1.81 &         &          &          &           & \underline {47}~~\underline {6.09}\\
1.59-2.51 & 1.15 (6) & 2.15 & ~6~~4.72& 29~~5.40 &  52~~5.66&  90~~5.89 &         & 10~~4.94 \\
2.00-2.51 & 7.66 (5) & 2.28 &         &          &          &           & 44~~5.76&          \\
2.51-3.16 & 1.53 (6) & 2.87 &         &          &          &  \underline {41}~~\underline {5.43} &         &          \\
2.51-3.98 & 4.58 (6) & 3.41 & 16~~4.54& 41~~4.95 &  72~~5.20&           & 18~~4.59&          \\
3.16-3.98 & 3.05 (6) & 3.62 &         &          &          &  26~~4.93 &         &          \\
3.98-5.01 & 6.08 (6) & 4.56 &         &          &  \underline {37}~~\underline {4.78}&           &         &          \\
3.98-6.31 & 1.82 (7) & 5.40 & \underline {18}~~\underline {3.99}& \underline {46}~~\underline {4.40} &  & ~5~~3.44& &\\
5.01-6.31 & 1.21 (7) & 5.73 &         &          &  32~~4.42&           &         &          \\
6.31-10.0 & 7.25 (7) & 8.55 & 11~~3.18& 29~~3.60 &  ~8~~3.04&           &         &          \\
~~~~$>$10.0 &          &    & ~1~~$--$&          &          &           &         &          \\
\hline
          &          &Total & 59~~~~~~~   & 173~~~~~~  & 233~~~~~~~& 253~~~~~~ & 154~~~~&76~~~~~~~\\
\hline
\end{tabular}
}  
\end{center}
\end{table*}

\begin{table}
\caption{The logarithmic space densities for late-type giants (symbols as 
in Table 2)}
\begin{center}
{\scriptsize
\begin{tabular}{rcccc}
\hline
$r_{1}-r_{2}$ &     $\Delta V_{1,2}$ &         $\bar{r}$ &   N &  D* \\
\hline
     ~~~~~0-3.98 &   6.11 (6) &      ~3.16 &         14 &       4.36 \\
     ~~3.98-6.31 &   1.82 (7) &      ~5.40 &         24 &       4.12 \\
     ~~6.31-7.94 &   2.42 (7) &      ~7.22 &         15 &       3.79 \\
     ~7.94-10.00 &   4.83 (7) &      ~9.09 &         18 &       3.57 \\
     10.00-12.59 &   9.64 (7) &      11.44 &         20 &       3.32 \\
     12.59-15.85 &   1.92 (8) &      14.40 &         20 &       3.02 \\
     15.85-19.95 &   3.84 (8) &      18.13 &         ~7 &       2.26 \\
     19.95-25.12 &   7.66 (8) &      22.83 &         ~8 &       2.02 \\
     25.12-31.62 &   1.53 (9) &      28.74 &         ~5 &       1.51 \\
     31.62-39.81 &   3.05 (9) &      36.18 &         ~5 &       1.21 \\
     ~~~$>$39.81 &            &            &         ~2 &       $--$ \\
\hline
\end{tabular}
}  
\end{center}
\end{table}

\begin{figure}
\resizebox{15cm}{!}{\includegraphics*{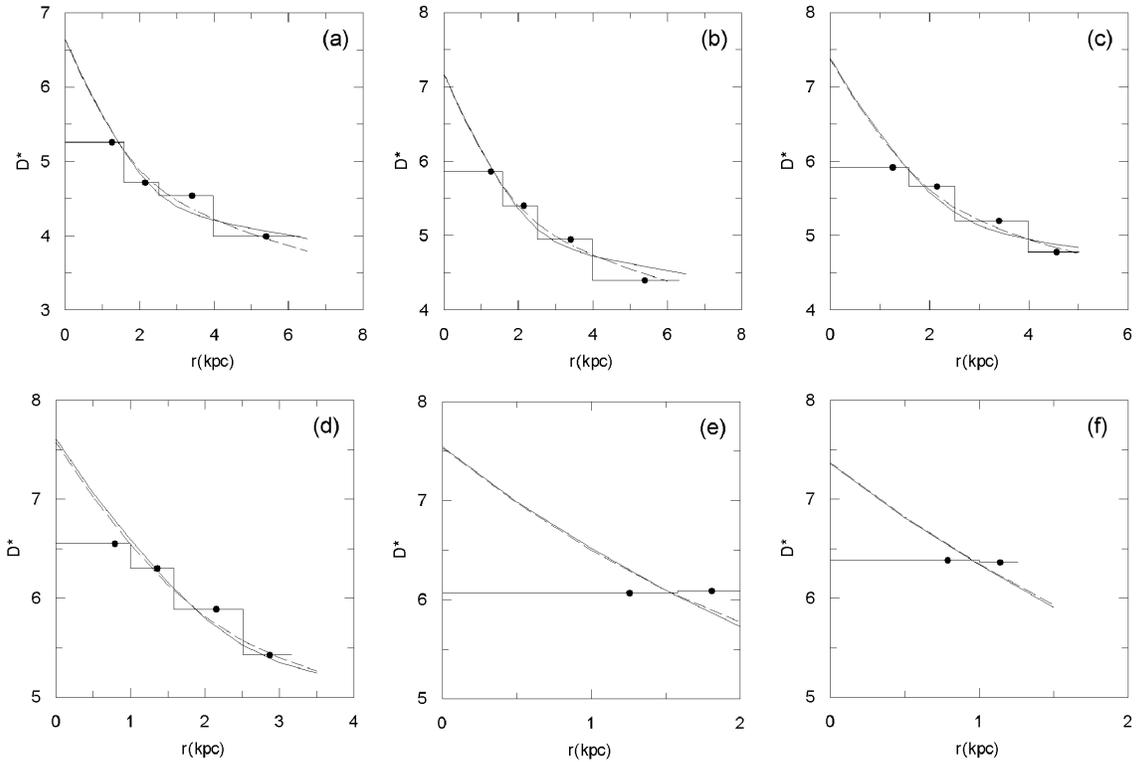}}
\caption{Logarithmic space-density histograms for all populations, within the 
limiting apparent magnitude of different $M(G)$ intervals: (a) $(2,3]$, 
(b) $(3,4]$, (c) $(4,5]$, (d) $(5,6]$, (e) $(6,7]$, and (f) $(7,8]$. ($\bullet$) 
centroid-distances within the limits of completeness for comparison with 
BRK (dashed curve) and GW (thin curve) model gradients.}
\end{figure}

\begin{figure}
\resizebox{15cm}{!}{\includegraphics*{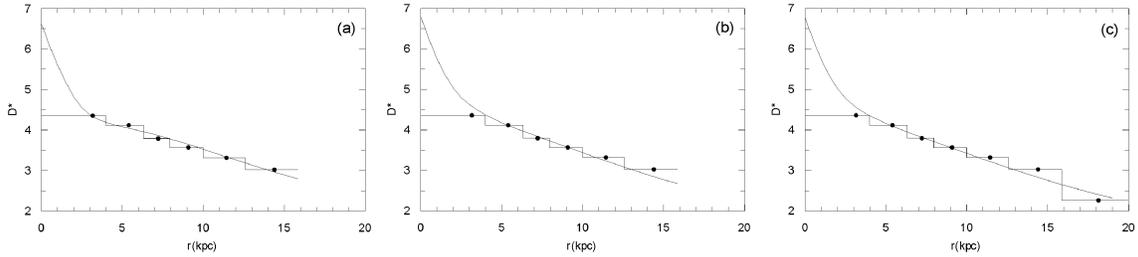}}
\caption{Logarithmic space-density histograms for giants for all populations 
($M(G)\leq2$ mag), within the limiting apparent magnitude. ($\bullet$) 
centroid-distance within the limiting distance of completeness for comparison 
with two model gradients, i.e. (a) GW, (b) BRK, and (c) BRK for a larger 
distance interval ($r\leq19.95$ kpc).}
\end{figure}

\begin{figure}
\resizebox{15cm}{!}{\includegraphics*{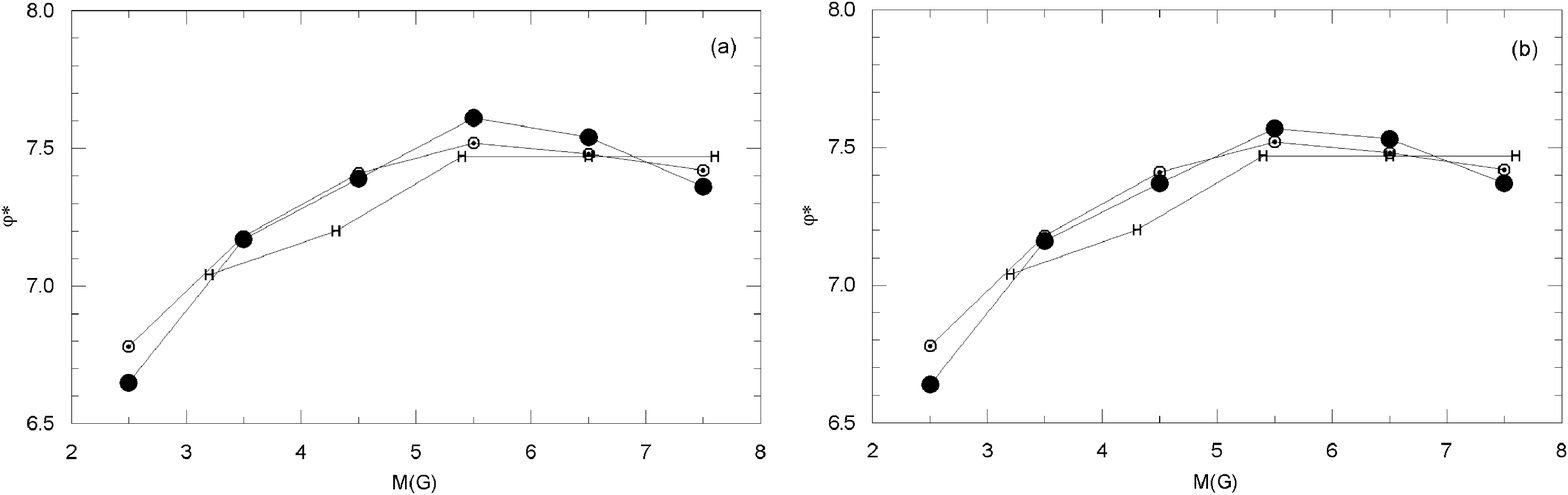}}
\caption{Logarithmic stellar luminosity function, implied by two 
model-gradients: (a) GW, and (b) BRK ($\bullet$) confronted to Gliese's 
(1969) ($\odot$), and Hipparcos' (JW 1997) (H) values.}
\end{figure}

\section{Metallicity}
The agreement of the observed space density functions with the model gradients,
and of the local densities with Gliese's (1969) and Hipparcos' (JW 1997) values 
confirms both the separation of the stars into different luminosity classes and 
their absolute magnitude determination. Then, we can use this advantage to 
investigate the metallicity distribution and clarify the question of a probable 
metallicity gradient in the direction to our field. As cited in Section 3, the 
new formula for the metallicity determination for dwarfs is valid throughout 
the interval $-2.20\leq[Fe/H]\leq+0.20$ dex, hence metal-abundances 
$[Fe/H]\leq-2.20$ or $[Fe/H]>+0.20$ evaluated by the same formula are less 
certain. However, the number of stars with extreme metal abundances is not 
large, especially the metal-poor ones; thus, they do not affect our results 
significantly.

The metallicity distribution for dwarfs peaks at $[Fe/H]\sim 0.1$ dex, show
a plateau between $[Fe/H]=-0.7$ and $-0.1$ dex, and decreases monotonously 
down to $[Fe/H]=-3$ dex (Fig.8a), however the metal-poor stars are small in
number. Hence, the intermediate and metal rich stars dominate the distribution.
Sub-giants, from the other hand, drawn from a larger spatial volume peak at a 
lower metallicity, i.e. $[Fe/H]=-0.6$ dex (Fig.8b).

The metallicities for all stars (dwarfs and sub-giants) summed over different 
$z$-distances show almost the same distribution, and hence a metallicity  
gradient can not be derived. On the other hand, dwarfs (only) with absolute 
magnitudes $4<M(G)\leq6$, corresponding to spectral types F5-K0, which are 
long lived main-sequence stars, do show different metallicity distributions and
reveal a metallicity gradient (Fig.9). Actually, the peaks for 
$[Fe/H]\cong-0.40$ and $\cong-0.80$ dex for the $z$-interval $0.75<z\leq1.5$ 
kpc in Fig.9b (marked with numbers 1 and 2), shift to $[Fe/H]\cong-0.60$ and 
$[Fe/H]\cong-1.00$ dex, respectively, for the interval $1.5 <z\leq2.5$ kpc in 
Fig.9c (again, marked with numbers 1 and 2). Thus, for both displacements we 
get $d[Fe/H]/dz\cong-0.20$ dex/kpc. No any radial metallicity gradient could 
be detected for the same sample.

\begin{figure}
\resizebox{15cm}{!}{\includegraphics*{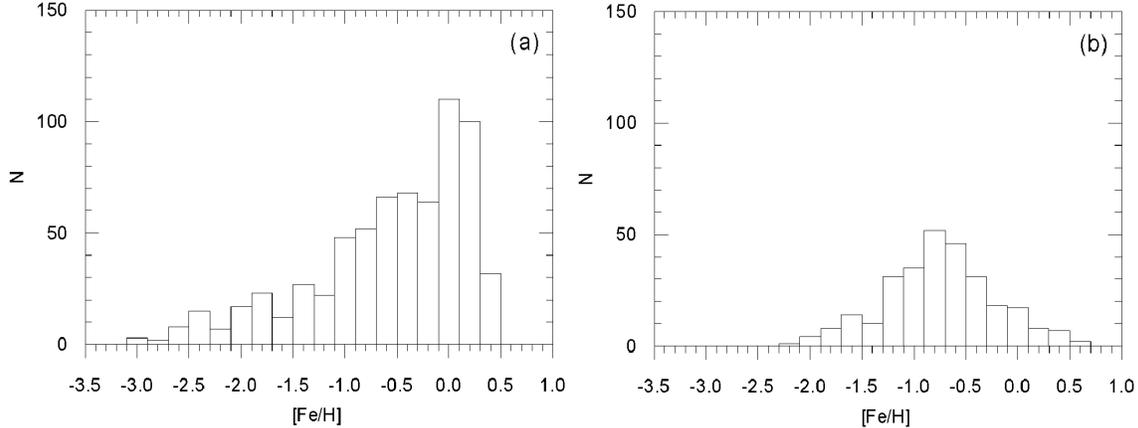}}
\caption{The metallicity distribution for dwarfs (a) and sub-giants (b).}
\end{figure}

\begin{figure}
\resizebox{15cm}{!}{\includegraphics*{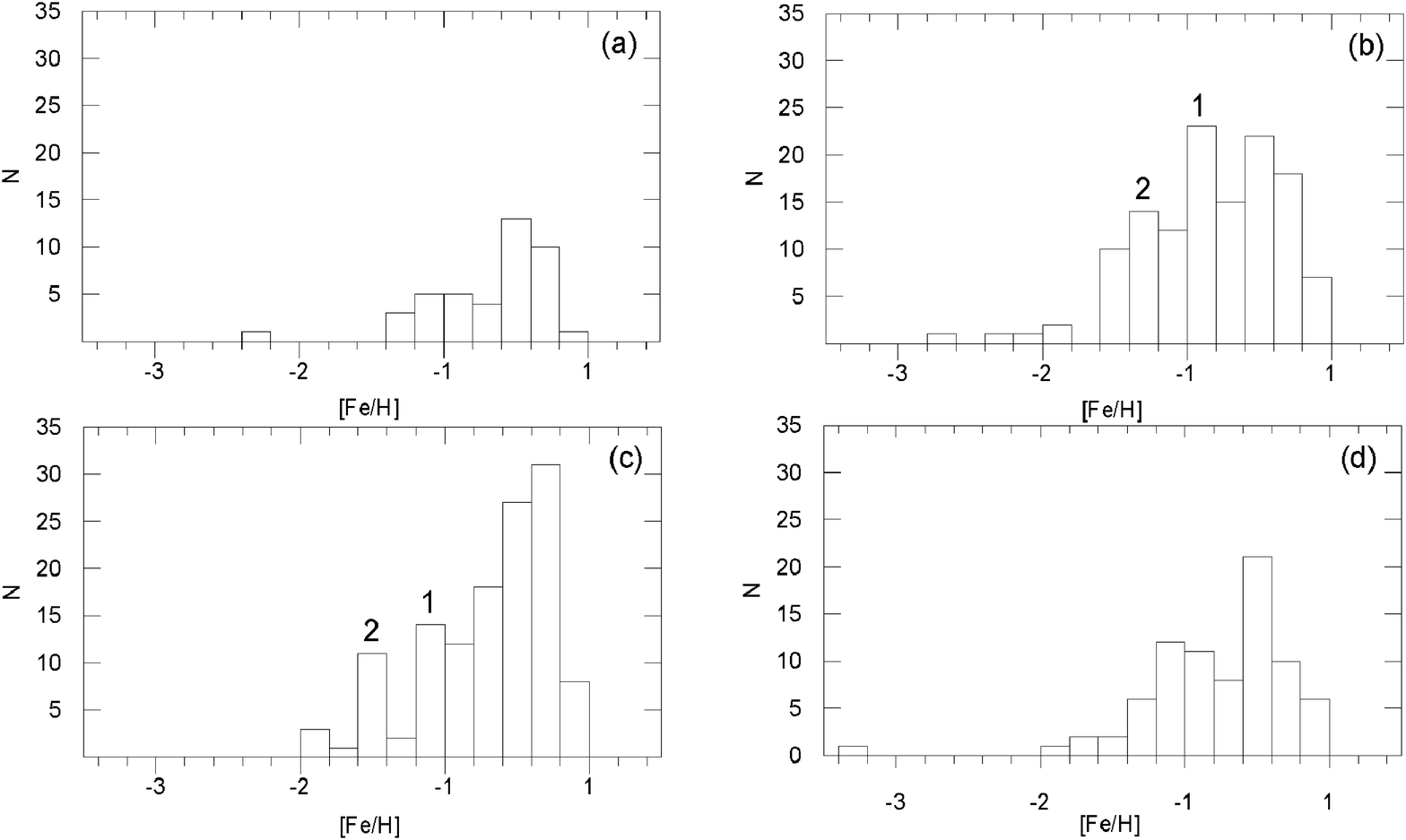}}
\caption{Metallicity distribution for dwarfs (only) with $4<M(G)\leq6$ mag, 
corresponding to spectral types F5-K0, for four $z$-intervals: (a) $0<z\leq0.75$, 
(b) $0.75<z\leq1.5$, (c) $1.5<z\leq2.5$, and (d) $2.5<z\leq4$ kpc. The comparison 
of Fig.9b and 9c reveals a metallicity gradient of $d[Fe/H]/dz=-0.20$ dex/kpc.}
\end{figure}

Within the limiting distance of completeness (arrows, Fig.10) the spatial 
distribution for dwarfs and sub-giants shows that $z=1.6$ kpc and $z=2.75$ kpc
are the borders of dominating regions of three populations, i.e. Population I 
(thin disk), Intermediate Population II (thick disk), and Extreme Population II
(halo) (for separation of field stars into different population types see 
Karaali 1994). Hence, the metallicity gradient cited above covers 
both disks. The metallicity distribution for dwarfs and sub-giants with 
$1.6<z\leq2.75$ kpc, i.e. for the thick disk, gives a bimodal distribution 
(Fig.11): the first mode, $[Fe/H]=-0.63$ dex, corresponds to the metal abundance 
assigned to the thick disk when it was introduced into the literature 
(Gilmore \& Wyse 1985, Wyse \& Gilmore 1986), and the second one, 
$[Fe/H]=+0.06$, to the metallicity which was cited for thick disk very recently 
(Carney 2000, Karaali et al. 2000). Although the number of super-solar metallicity
stars seems to be larger than usually expected at distance $z=1.6$ kpc above the
galactic plane, we accept this result regarding the procedures used for distance 
and metallicity evaluation. Actually, the luminosity functions implied by two model
gradients cited above (Fig.7) show that our distance estimation is accurate, 
as well as the procedure of Carney (1979) (see Appendix) adopted for metallicity 
estimation.      

\begin{figure}
\resizebox{15cm}{!}{\includegraphics*{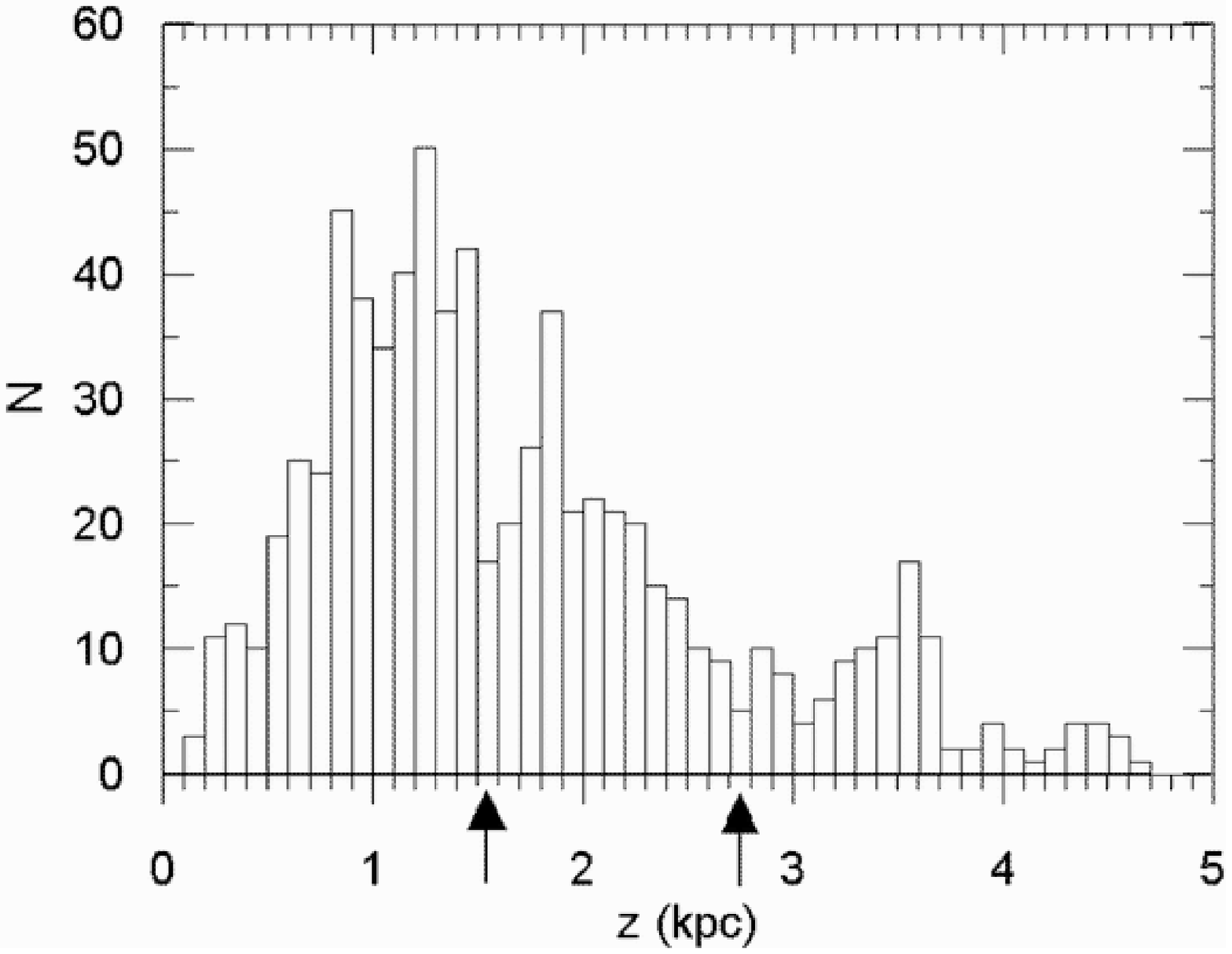}}
\caption{The spatial distribution for dwarfs and sub-giants within the 
limiting distance of completeness. The distances to the Galactic plane $z=1.6$ 
kpc and $z=2.75$ kpc (shown by arrows) are the borders of three populations, 
Population I (thin disk), Intermediate Population II (thick disk), and Extreme 
Population II (halo).}
\end{figure}
 
\begin{figure}
\resizebox{15cm}{!}{\includegraphics*{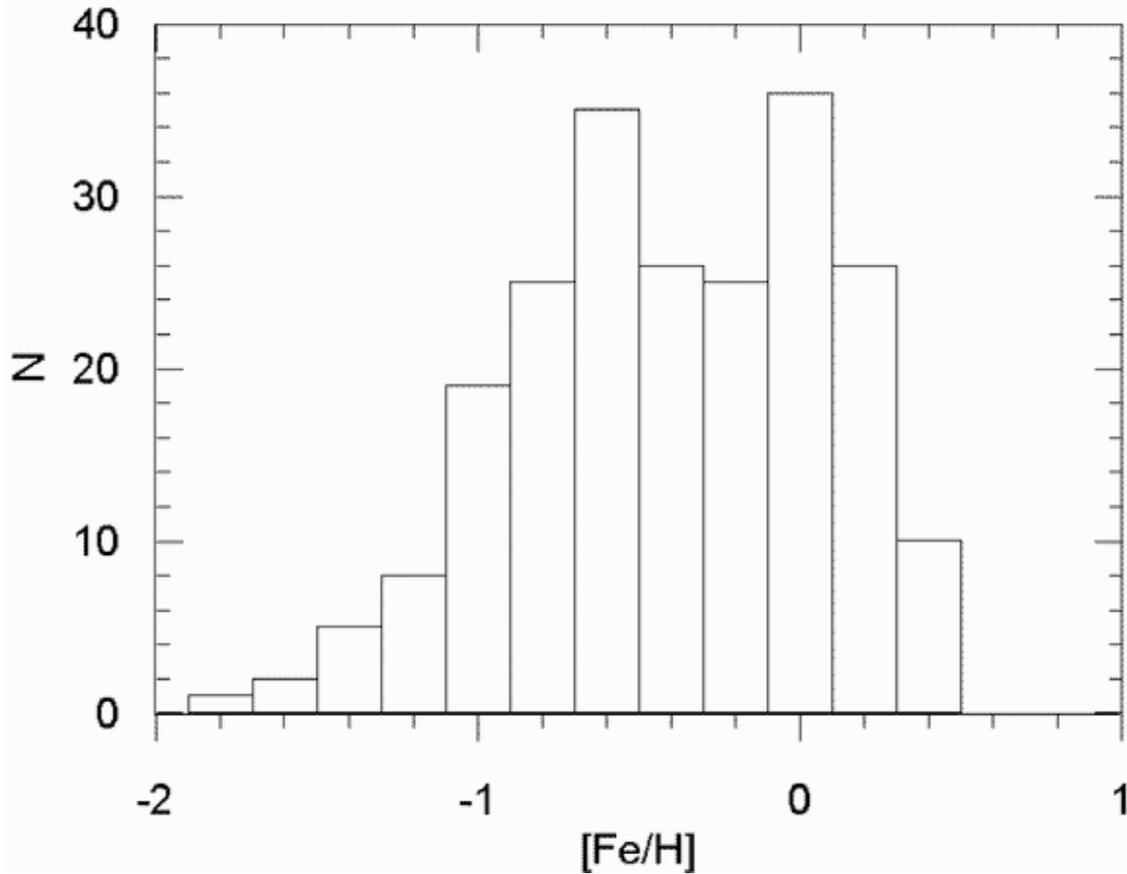}}
\caption{The metallicity distribution for dwarfs and sub-giants with 
$1.6<z\leq2.75$ kpc, i.e. for the thick disk, giving a bimodal distribution. 
The first mode at $[Fe/H]=-0.63$ dex corresponds to the canonical metal
abundance assigned to the thick disk (Gilmore \& Wyse 1985, Wyse \& Gilmore 1986, 
Buser et al. 1999, and Rong et al. 2001), and the second at $[Fe/H]=+0.06$ dex to 
the metallicity cited for the same component of the Galaxy very recently 
(Carney 2000, Karaali et al. 2000).}
\end{figure}

\section {Summary and conclusion}

The re-investigation of this field has been attractive because it provides 
illustrative and effective applications of a number of tools: the 
extra-galactic objects identified by comparison of Basel and Minnesota charts, 
and a small region with size 0.104 square-degrees contaminated by stars of 
the cluster M5 and affected by background light of bright star HD 136202  
($\alpha = 15^{h}~ 19^{m}~18^{s}.80$, $\delta=+01^{o}~45{'}~55{''}.5$, epoch 
2000) are excluded from the program. Additionally, the evolved stars 
(sub-giants and giants) are separated from the dwarfs by an empirical method 
which has been used successfully in our recent works (Karaali 1992, Karaali 
et al. 1997, Ak et al. 1998, Karata\c{s} et al. 2001, Karaali et al. 2003), and 
the absolute magnitudes are determined by the colour-magnitude diagrams of 
Buser \& Fenkart (1990), and Buser et al. (2000), obtained via synthetic 
photometry. Resulting logarithmic space densities agree with the model gradients 
of BRK, and GW. The luminosity function, which reflects the local densities, 
also agrees with the Gliese (1969) and Hipparcos (JW 1997) functions for the 
absolute magnitude intervals $2<M(G)\leq3$, $3<M(G)\leq4$, $4<M(G)\leq5$, 
$5<M(G)\leq6$, $6<M(G)\leq7$, and $7<M(G)\leq8$. However, for the giants two 
different local densities are in consideration, i.e. the comparison of observed 
space density histograms with the model gradients of GW give $D^{*}(0)=6.63$, 
rather close to that of Gliese (1969), and a slightly higher value, 
$D^{*}(0)=6.79$, when compared with BRK model gradients. From this we can 
conclude that the separation of field stars into different categories is 
probably carried out correctly. 

The agreement cited above is used to advantage in treating the metallicity 
distribution for dwarfs and subgiants, and in looking for a probable 
metallicity gradient in this direction of the Galaxy. No different distribution 
can be observed for different $z$-distances from the Galactic plane, when 
dwarfs and sub-giants are considered together. However, this is not the case 
for dwarfs only, with absolute magnitudes $4<M(G)\leq6$, corresponding to 
spectral types F5-K0, the long lived main-sequence stars. The difference 
between the two peaks in the metallicity distribution for $0.75<z\leq1.5$ 
kpc and $1.5<z\leq2.5$ kpc reveals a metallicity gradient 
$d[Fe/H]/dz\cong-0.20$ dex/kpc (Fig.9). 

The metallicity for the thick disk ($1.6<z\leq2.75$ kpc) shows a bimodal  
distribution (Fig.11): the first mode, $[Fe/H]=-0.63$ dex, represents the 
canonical metal abundance assigned to the thick disk (Gilmore \& Wyse 
1985, Wyse \& Gilmore 1986, Buser et al. 1999, Rong et al. 2001, 
Karaali et al. 2003), and the second, $[Fe/H]=+0.06$ dex, corresponds to the 
value cited very recently (Carney 2000, Karaali et al. 2000). The metal-poor 
tail claimed by many authors (Rogers \& Roberts 1993, Layden 1995, 
Beers \& Sommer-Larsen 1995, Norris 1996, Chiba \& Yoshii 1998, and 
Karaali et al. 2000) also exists in this direction.
  
This is one of the individual-field investigations of the Basel program, which 
offers space density functions in agreement with two Galactic model gradients, 
local space densities close to Gliese's (1969) and Hipparcos' (JW 1997) values, 
and vertical metallicity gradients which cover both the thin and thick disks. 
Thus, we confirm the works of Reid \& Majewski (1993), Chiba \& Yoshii (1998), 
Buser et al. (1999), Rong et al. (2001), and Karaali et al. (2003).

\section*{Appendix}
We adopted the procedure used by Carney to obtain an equation for deriving the 
metallicity of a dwarf star from its observed ultra-violet excess, 
$\delta_{U-G}$. Two steps were followed for our purpose: in the first step, 
$UBV$ data for 52 and 24 dwarfs taken from Carney (1979) and Cayrel de Strobel 
et al. (1997), respectively, are transformed to the $RGU$ system by means of 
the metallicity-dependent conversion equations of G\"{u}ng\"{o}r-Ak (1995). 
The ($U-G$, $G-R$) main-sequence of the Hyades, transformed from $UBV$ to $RGU$ 
by the same formulae, is used as a standard sequence for ultra-violet excess 
evaluation. The transformation formulae just cited, or those of Buser (1978), 
may be used to show that the {\em guillotine factors\/} given by Sandage (1969) 
for $UBV$ photometry also apply for normalizing the ultra-violet excesses obtained
on the $RGU$-photometric system, as follows: The equations which transform 
$U-B$ and $B-V$ colour indices of a star to the $G-R$ and $U-G$ colour indices 
are generally given by 
\begin{equation}
G-R = a_{1}(U-B) + b_{1}(B-V) + c_{1}, 
\end{equation}

\begin{equation}
U-G = a_{2}(U-B) + b_{2}(B-V) + c_{2}, 
\end{equation}
where $a_{i}$, $b_{i}$, and $c_{i}$ (i = 1, 2) are parameters to be determined.
Let us write equation (2) for two stars with the same $B-V$ (or equivalently 
$G-R$), i.e. for a Hyades star (H) and for a star (*) whose ultra-violet 
excess would be normalized,
\begin{equation}
(U-G)_{H} = a_{2}(U-B)_{H} + b_{2}(B-V) + c_{2},
\end{equation}
                             
\begin{equation}
(U-G)_{*} = a_{2}(U-B)_{*} + b_{2}(B-V) + c_{2}.
\end{equation}
Then, the ultra-violet excess for the star in question, relative to the Hyades 
star is,
\begin{equation}
(U-G)_{H} - (U-G)_{*} = a_{2}[(U-B)_{H} - (U-B)_{*}]
\end{equation}
or, in standard notation,
\begin{equation}
\delta(U-G) = a_{2} \delta(U-B).
\end{equation}
Now, for another star with the same metal-abundance $[Fe/H]$ but with 
$B-V=0.60$ mag, (or its equivalent $G-R=1.08$) we get, in the same way,
\begin{equation}
\delta (U-G)_{1.08} = a_{2} \delta(U-B)_{0.60}.
\end{equation}
Equations (6) and (7) give,

\begin{equation}
\frac{\delta(U-G)_{1.08}}{\delta(U-G)}=\frac{\delta(U-B)_{0.60}}{\delta(U-B)}=f,
\end{equation}
where $f$ is the ultra-violet excess conversion (or {\em guillotine\/}) factor 
in question. Hence, the $RGU$-photometric $\delta(U-G)$ can be normalised by 
the same $f$ factors as are used in $UBV$ photometry. 

In the second step, 76 stars are separated into 14 metallicity intervals, with 
different bin sizes, chosen such as to provide an almost equal number of stars 
in each bin. The least-squares method is used to obtain a calibration between 
the normalized ultra-violet excess $\delta_{1.08}$ and metallicity $[Fe/H]$. 
This binning provides equal-weight data for 14 points in Fig.12, which 
represent the mean metallicities and mean $\delta_{1.08}$ excesses for each 
bin. The constant term $a_{o}$ in the equation,

\begin{equation}
[Fe/H] = a_{o} + a_{1}\delta_{1.08} + a_{2}\delta^{2}_{1.08}
\end{equation}
is assumed to be $a_{o}=0.11$ for consistency with the metallicity of the 
Hyades cited by Carney (1979). The least-squares method gives  
$a_{1}=-2.22$ and $a_{2}=-7.95$; thus,

\begin{equation}
[Fe/H] = 0.11 - 2.22\delta_{1.08} - 7.95\delta^{2}_{1.08}.
\end{equation}

\begin{figure}
\resizebox{15cm}{!}{\includegraphics*{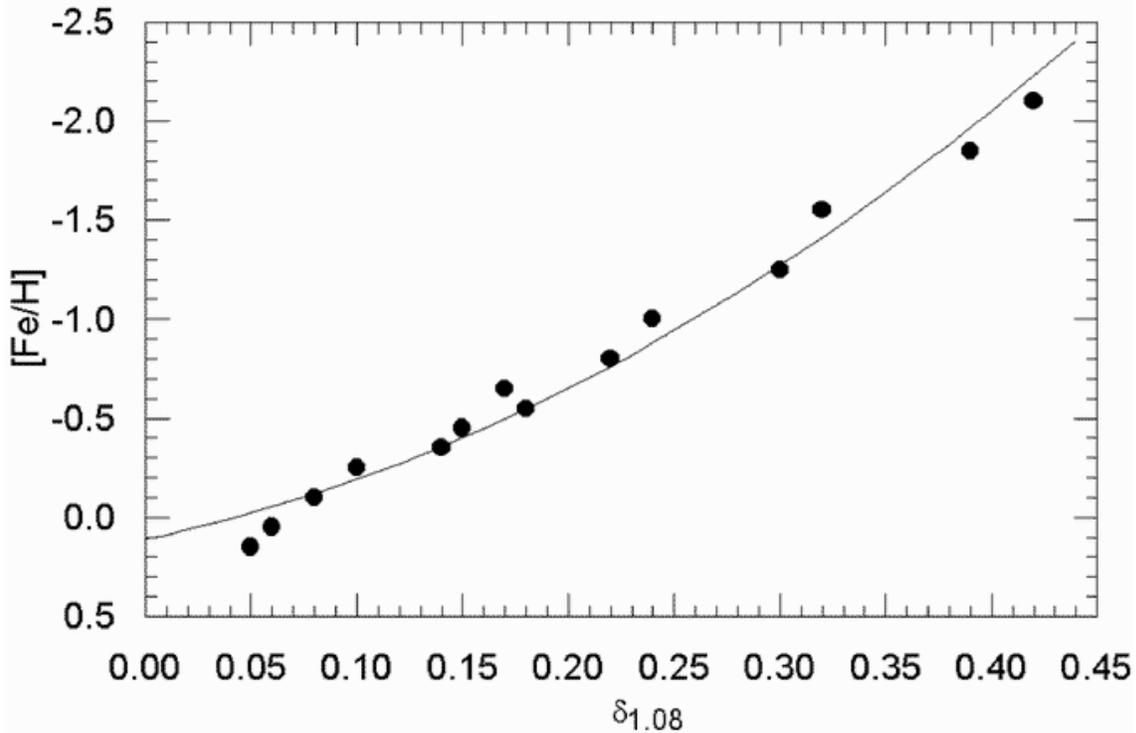}}
\caption{$[Fe/H]$ metallicity versus normalized $\delta_{1.08}$ ultra-violet 
excess for $RGU$ photometry.}
\end{figure}

\begin{figure}
\resizebox{15cm}{!}{\includegraphics*{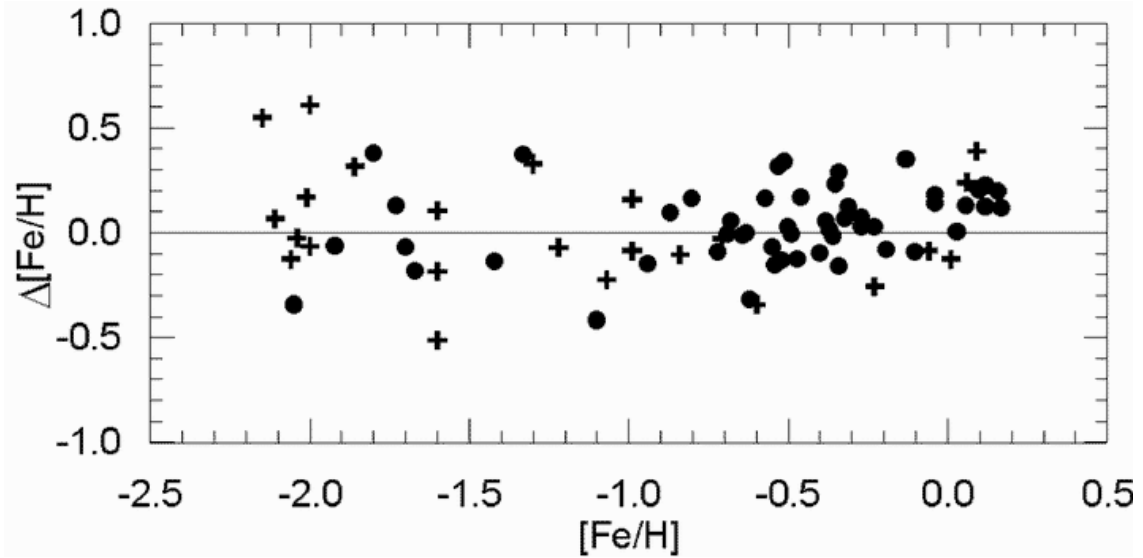}}
\caption{$\Delta [Fe/H]$ versus metallicity, where $\Delta [Fe/H]$ is the 
difference between the original metallicities and the evaluated ones, utilising
the new calibration, $[Fe/H]=0.11-2.22\delta_{1.08}-7.95\delta^{2}_{1.08}$. 
Symbols: ($\bullet$) stars from Carney (1979), and (+) stars from 
Cayrel de Strobel et al. (1997).}
\end{figure}

The differences between the metallicities evaluated by means of equation (10) 
and the original ones, i.e. $\Delta[Fe/H]$, versus the original metallicities 
are given in Fig.13. The differences are large only for a few metal-poor stars,
while the scatter relative to the line $\Delta[Fe/H]=0.0$ dex is small. 
Actually the mean of the differences (for all stars) is only 0.02 dex, while 
the probable error for the mean is small, $p.e.= \pm 0.15$ dex, indicating that
the new calibration can be used with good accuracy.

Dwarfs used for the new metallicity calibration are identified either according
to their spectral types or surface gravities.

\section*{Acknowledgments}
We acknowledge financial support by Research Fund of the Istanbul University 
through Project No: \"{O}-970/23032001. Also we would like to thank the Swiss 
National Science Foundation for financial support, and to the University of 
Minnesota for providing us APS on-line catalogues for the field investigated 
in this study.

\section*{References}

\reference Ak, S. G., Karaali, S., \& Buser, R. 1998, A\&AS, 131, 345

\reference Becker, W., Steinlin, U., \& Wiedeman, D. 1978, Photometric 
catalogue for stars in selected areas and other fields in the $RGU$-system 
(V). Basel; Astronomical Institute of the University of Basel

\reference Beers, T. C., \& Sommer-Larsen, J. 1995, ApJS, 96, 175

\reference Bilir, S., Karata\c{s}, Y., \& Ak, S. G. 2003, TJPh, 27, 235

\reference Buser, R. 1978, A\&A, 62, 425

\reference Buser, R., \& Fenkart, R. P. 1990, A\&A, 239, 243

\reference Buser, R., \& Rong, J. 1995, BaltA, 4, 1

\reference Buser, R., Rong, J., \& Karaali, S. 1998, A\&A, 331, 934 (BRK)

\reference Buser, R., Rong, J., \& Karaali, S. 1999, A\&A, 348, 98

\reference Buser, R., Karata\c{s}, Y., Lejeune, Th., Rong, J. X., Westera, 
P., \& Ak, S. G. 2000, A\&A, 357, 988

\reference Carney, B. W. 1979, ApJ, 233, 211

\reference Carney, B. W. 2000, Proceedings of the $35^{th}$ Li\'ege Int. 
Astrophys. Coll., July 5-8, 1999, p. 287

\reference Cayrel de Strobel, G., Soubiran, C., Friel, E. D., Ralite, N., 
\& Francois, P. 1997, A\&AS, 124, 299

\reference Chiba, M., \& Yoshii, Y. 1998, AJ, 115, 168

\reference Del Rio, G., \& Fenkart, R. P. 1987, A\&AS,  68, 397

\reference Fenkart, R. P., \& Karaali, S. 1987, A\&AS,  69, 33

\reference Fenkart, R. P., \& Karaali, S. 1990, A\&AS,  83, 481

\reference Gilmore, G., \& Wyse, R. F. G. 1985, AJ, 90, 2015 (GW)

\reference Gliese, W. 1969, Veröff. Astron. Rechen Inst. Heidelberg No:22

\reference G\"{u}ng\"{o}r-Ak, S. 1995, Ph. D. Thesis, Istanbul Univ.

\reference Jahreiss, H., \& Wielen, R. 1997, in: HIPPARCOS'97. Presentation 
of the HIPPARCOS and TYCHO catalogues and first astrophysical results of the 
ipparcos space astrometry mission., Battrick, B., Perryman, M. A. C., \& 
ernacca, P. L., (eds.), ESA SP-402, Noordw\"{y}k, p.675 (JW)

\reference Karaali, S. 1992, VIII. Nat. Astron. Symp. Eds. Z. Aslan and 
O. G\"{o}lba\c{s}\i, Ankara - Turkey, p.202

\reference Karaali, S. 1994, A\&AS, 106, 107

\reference Karaali, S., Karata\c{s}, Y., Bilir, S., \& Ak,  S. G. 1997, 
IAU Abstract Book, Kyoto, p.299

\reference Karaali, S., Karata\c{s}, Y., Bilir, S., Ak, S. G., \& Gilmore, 
G. F. 2000, The Galactic Halo: from Globular Clusters to Field Stars. 
$35^{th}$ Li\'ege Intr. Astr. Coll. Edts. A. Noel et al., p. 353

\reference Karaali, S., \& Bilir, S. 2002, TJPh, 26, 427

\reference Karaali, S., Ak, S. G., Bilir, S., Karata\c{s}, Y. \& Gilmore, G. 
2003, MNRAS, 343, 1013 

\reference Karata\c{s}, Y., Karaali, S., \& Buser, R. 2001, A\&A, 373, 895

\reference Layden, A. C. 1995, AJ, 110, 2288

\reference Norris, J. E. 1996, ASP Conf. Ser. Vol.92, p.14, Eds. H. L. 
Morrison \& A. Sarajedini

\reference Reid, N., \& Majewski, S. R. 1993, ApJ, 409, 635

\reference Rodgers, A. W., \& Roberts, W. H. 1993, AJ, 106, 1839

\reference Rong, J., Buser, R., \& Karaali, S. 2001, A\&A, 365, 431

\reference Sandage, A. R. 1969, ApJ, 158, 1115

\reference Schlegel, D. J., Finkbeiner, D. P., \& Davis, M. 1998, ApJ, 
500, 525

\reference Wyse, R. F. G., Gilmore, G. 1986, AJ, 91, 855

\end{document}